\documentclass{ifacconf}

\usepackage{graphicx}      
\usepackage{natbib}        
\usepackage[ansinew]{inputenc}
\usepackage{amsmath}
\usepackage{subfig}
\usepackage{epstopdf}
\usepackage{bm}
\usepackage{dblfloatfix}
\usepackage{tabulary}
\usepackage{tabularx}
\usepackage{booktabs}
\usepackage{scrextend}
\usepackage{amssymb}
\usepackage[width=\textwidth]{caption}

\begin{document}
\begin{frontmatter}
\title{Security Solutions for Local Wireless Networks in Control Applications based on Physical Layer Security} 

\thanks[footnoteinfo]{This is a preprint, the full paper has been published in Proceedings 
of the 3rd IFAC Conference on Embedded Systems, 
Computational Intelligence and Telematics in Control - CESCIT 2018, \copyright 2018 IFAC.}

\author{Andreas Weinand, Michael Karrenbauer, Hans D. Schotten} 

\address{Institute for Wireless Communication and Navigation\\University of Kaiserslautern, Germany\\e-mail: \{weinand, karrenbauer, schotten\}@eit.uni-kl.de}

\begin{abstract}                
The Design of new wireless communication systems for industrial applications, e.g. control applications, is currently a hot research topic, as they deal as a key enabler for more flexible solutions at a lower cost compared to systems based on wired communication. However, one of their main drawbacks is, that they provide a huge potential for miscellaneous cyber attacks due to the open nature of the wireless channel in combination with the huge economic potential they are able to provide. Therefore, security measures need to be taken into account for the design of such systems. Within this work, an approach for the security architecture of local wireless systems with respect to the needs of control applications is presented and discussed. Further, new security solutions based on Physical Layer Security are introduced in order to overcome the drawbacks of state of the art security technologies within that scope. 
\end{abstract}


\end{frontmatter}
\onecolumn

\section{Introduction}
\label{intro}
    
A current trend in industrial environments is the utilization of wireless systems in order to serve for applications such as metering, closed loop control processes or remote maintenance of machines or even whole plants without having the drawbacks of wired communication systems. Additionally, the availability of such machine-to-machine (M2M) communication systems may even lead to completely new applications in future. Especially due to their high flexibility, wireless systems can enable the connection of a huge number of devices in such an environment at a low cost. Typically, applications in industrial scenarios have high requirements regarding message latency and reliability. This type of traffic is also referred to as mission critical machine type communication (MC-MTC) or ultra-reliable low latency communication (URLLC) [\cite{ITU-R2015}]. Another important requirement within industrial applications is to guarantee secure exchange of data over the wireless network in order to prevent cyber attacks.\\
Due to sensitive information, such as process control data or machine configuration files, which is transferred over the wireless interface in industrial applications, every possible participant within the coverage of the network is able to interact with the transmitted information and consequently able to interact with the served applications as well. Therefore, dedicated security measures in order to protect this sensitive information are needed. Otherwise the safe operation of the controlled machines and processes can not be guaranteed and additionally the disclosure of intellectual property such as process control parameters, machine configuration data or even simple information such as the production volume may leak. Possible attacks that might prohibit the secure transfer of sensitive data on wireless interfaces include passive attacks such as eavesdropping, as well as active attacks such as replay attacks, man-in-the-middle attacks or address spoofing attacks. In general, the wireless channel is at least insecure during the initial communication phases. In order to establish a secured wireless link between two or more network devices, it has to be guaranteed that initial trust is set up between them during that stage. Additionally, there are two more goals that need to be achieved by security measures. These are prevention of data manipulation during transmission over the air which can be caused by launching active attacks and secondly protection of data confidentiality which can be achieved by eavesdropping on ongoing transmissions. Altogether, the following security requirements can be derived:
\begin{enumerate}
	\item Ensure confidentiality, integrity and authenticity of messages on the wireless interface of the system
	\item Allow only desired devices access to the network
	\item Ensure life cycle security of network devices
	\item High degree of usability 
\end{enumerate}
There are also other requirements influencing the design of the security architecture and components of wireless systems supporting URLLC. Especially, the system should be flexible in such a sense, that it is easily possible to add new devices to an existing network or replace network devices without huge configuration efforts. Also the installation of a whole new network should be possible with high usability for non experts (no manual configuration required by e.g. security engineers). In order to understand what implications the different security solutions have on URLLC, it is necessary to understand the main chracteristics of this type of traffic first. URLLC traffic has two key requirements as the name already implies, one is the reliability requirement and the other one is the latency requirement. As typical applications for URLLC are e.g. closed loop control processes, both of these requirements need to be guaranteed jointly, which means that messages have to be delivered with a given packet reception ratio (PRR) within a given time. If the latency requirement is not met, the message is considered as erroneous because its content is outdated in the sense of the application. Another key characteristic of URLLC is the message frequency or periodicity which is depending on the required update time of the control application and can be up to $1000$ Hz in case of very fast closed loop processes [\cite{Bockelmann2017}]. Further, also the requirements for the message payload size are different compared to conventional applications such as e.g. media streaming. URLLC systems need especially support for very short messages (down to a few Bytes [\cite{Bockelmann2017}]) which can e.g. consist of a single sensor value or a simple binary control command.\\
The remainder of the work is organized as follows. Section \ref{SotA} gives an overview of the State of the Art security solutions in local wireless networks. In section \ref{system_model} we present the system model which deals as a baseline for our work. Section \ref{urllc_security} introduces the key enabling technologies which are proposed in order to fulfill the mentioned security requirements and in section \ref{physec_section} we propose enhancements for these solutions based on physical layer security. Finally, section \ref{CONC} concludes the paper. 

\section{State of the Art}
\label{SotA}
In this section, current wireless protocols which are widely used for M2M communication are analyzed with respect to their security features. These are device authentication, key management and data confidentiality, integrity and authenticity. The most prominent networks which are currently used in the field of industrial applications are IEEE 802.11 based networks (WiFi). The protocol specification including security aspects for these kind of networks can be found in [\cite{802_11_standard}]. Regarding the device authentication, two options are possible. Either a pre-shared key (PSK) is used, which has to be known by both devices authenticating mutually and typically requires a user interaction, or a certificate based authentication process is used which is based on the IEEE 802.1X protocol [\cite{802_1X_standard}] and basically makes use of the Extensible Authentication Protocol (EAP) and X.509 based certificates. Pairwise session keys as well as group keys are derived from the initial authentication credentials (either from the PSK based credentials or from EAP parameters). The Counter Mode Cipher Block Chaining Message Authentication Code (CBC-MAC) Protocol (CCMP) is used in order to ensure message confidentiality (based on the AES block cipher [\cite{AES_standard}]) as well as message authenticity and integrity, based on Message Integrity Codes (MIC). In IEEE 802.15.1 based networks [\cite{802_15_1_standard}] such as Bluetooth [\cite{bluetooth}], the device authentication between $2$ devices as well as the following process of key derivation is based on a PIN, which needs to be entered via user interaction to a device. In earlier versions of the specification, only at least $4$ alphanumeric characters were required. Additionally, the proposed ciphers for providing message confidentiality (E0 stream cipher) as well as authenticity/integrity (SAFER+ cipher) in earlier versions are not considered as secure anymore nowadays. Another type of networks widely used in industrial applications are IEEE 802.15.4 based networks [\cite{802_15_4_standard}] such as e.g. Zigbee [\cite{zigbee}]. They do not provide any kind of certificate based device authentication (instead shared symmetric keys are used for that). Also the key management and key derivation is based on the initial keys, but authenticated encryption based on AES and Message Integrity Codes (MIC) is provided in order to guarantee message confidentiality, integrity and authenticity. In summary, IEEE 802.11 based wireless networks already provide the necessary means in order to guarantee the security requirements mentioned before. Within this work, we additionally investigate the effects of these security features on the system design in combination with URLLC specific aspects and propose solutions which adapt to the requirements of this type of traffic.

\section{System Model}
\label{system_model}
In this section we present the considered system model with respect to the requirements of industrial applications. We further introduce assumptions that are made within the scope of our work as well as the considered attacker model.

\begin{figure}[h]
\begin{center}
\includegraphics[width=8.4cm]{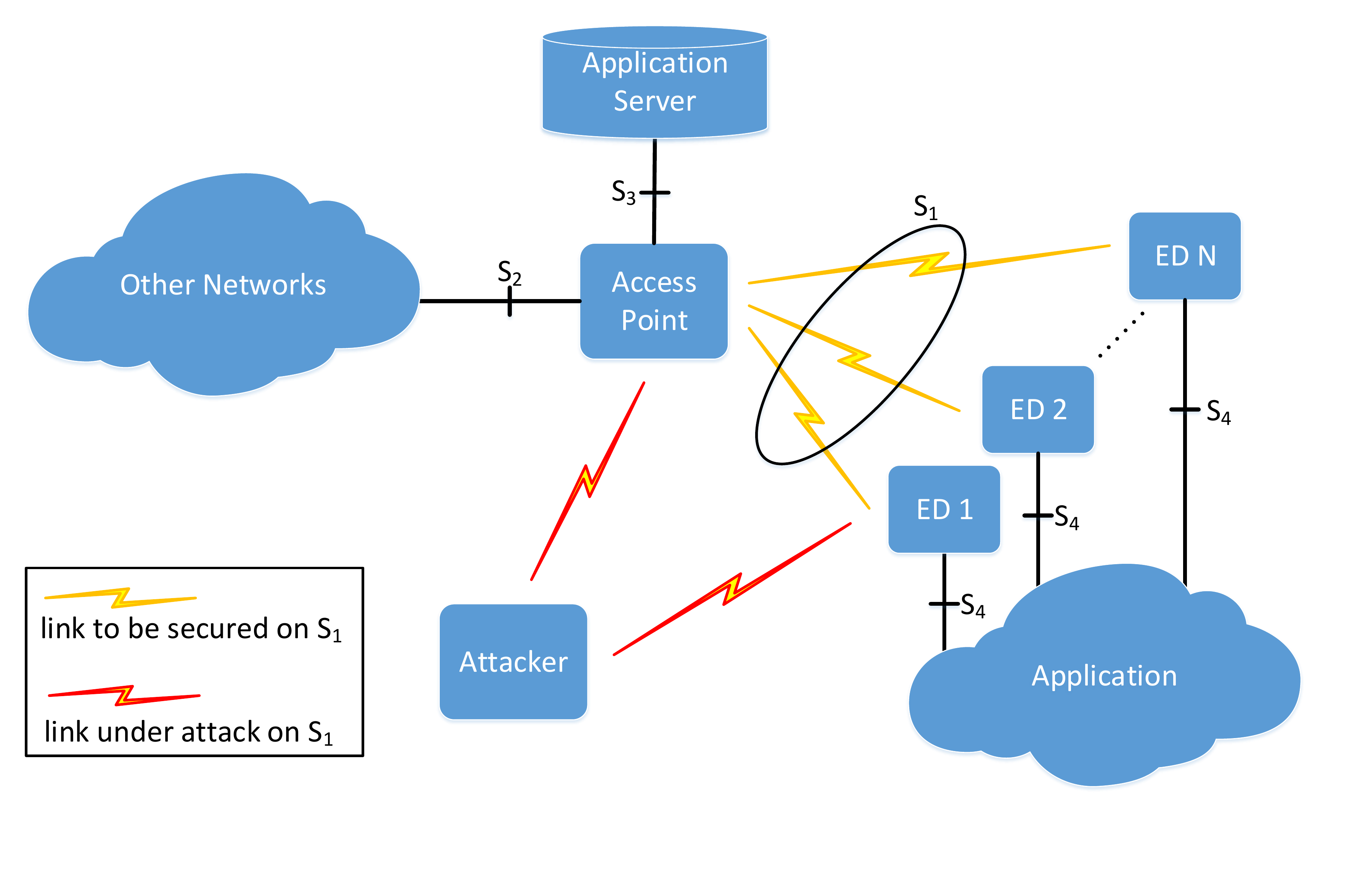}    
\caption{System Level Architecture} 
\label{sys_arch}
\end{center}
\end{figure}

\subsection{System Level Architecture}
Fig. \ref{sys_arch} shows the system level architecture which is consisting of network components and an application. The network devices are the Access Point (AP) and End Devices (ED) which both share the wireless interface $S_1$ (bidirectional). The application consists of an application server and the application clients (e.g. sensors and/or actuators). The interface $S_3$ is used in order to connect the application server with the AP, while interface $S_4$ is shared by the EDs and the application clients. With these interfaces, the application server can send control information to the application clients (actuators) and receive control update information vice versa. An additional interface $S_2$ is introduced in order to enable features such as e.g. human machine interaction (HMI) for system diagnosis purposes or getting access to further enterprise back-ends such as authentication servers.

\subsection{Assumptions}
The goal of our work is to propose solutions in order to secure the communication on the wireless interface $S_1$. Therefore, we assume that on the other interfaces $S_2$, $S_3$ and $S_4$ secure transmission links are already set up. Additionally, we assume that it is possible to store security credentials, which are necessary on order to set up a secure transmission links on $S_1$, in physically secure and tamper-proof storage such as secure elements (SE) for every network device (AP and EDs). Further, the SE is assumed to provide at least the following features:
\begin{enumerate}
	\item Secure storage, generation and verification of certificates
	\item Secure storage and generation of asymmetric key pairs
	\item Secure storage and generation of symmetric key pairs
\end{enumerate}
In our work we also assume, that network devices are already equipped with the above mentioned credentials. The process of loading these into the respective device (actually to its SE) is not scope of this work and approaches on how to achieve that goal can e.g. be found in [\cite{Mackenthun.2017}].

\subsection{Attacker Model}
A typical scenario for an attacker within the scope of our work is that he is located outside of a factory using advanced equipment such as directed antennas and high sensitivity receivers in order to maximize the communication range to his benefit. It is further assumed, that he has full knowledge of the used communication protocols in order to interact with the system over the wireless interface $S_1$ as shown in fig. \ref{sys_arch}. Thereby the attacker may either act as an AP and eavesdrop on uplink messages sent from EDs to the dedicated AP and spoof downlink messages to EDs in order to manipulate the application or he acts as an ED and spoofs messages to the AP in order to manipulate the control application on the application server and eavesdrop on downlink messages sent from the AP to the dedicated EDs. We do not consider attacks regarding the device authentication process which could be used by an attacker to gain access to the network. 

\section{Security for URLLC}
\label{urllc_security}
In this section, we introduce the key enabling components which are necessary in order to fulfill the given security requirements. We show their advantages and especially disadvantages in combination with the URLLC specific requirements. Alternative solutions in order to overcome these drawbacks, especially based on physical layer security, will then be presented within section \ref{physec_section}.

\begin{figure}[t]
\begin{center}
\includegraphics[width=6.3cm]{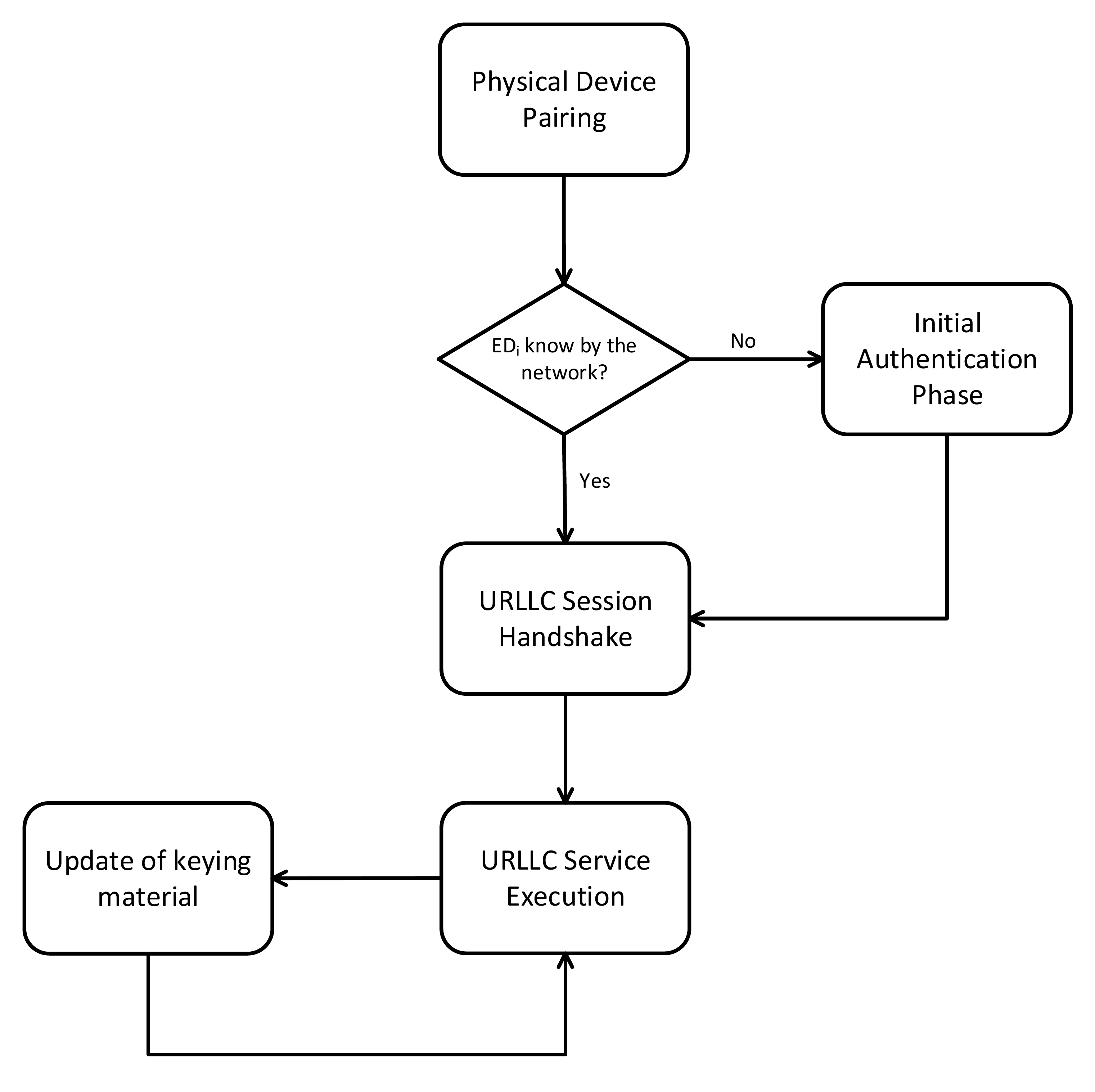}    
\caption{Overview of Plug\&Trust Procedures} 
\label{plug+trust_logic}
\end{center}
\end{figure}

\subsection{Plug\&Trust Protocols}
In order to guarantee secure communication on the wireless interface $S_1$, network devices (e.g. an AP and ED pair) need to set up trust initially towards each other. Therefore, basically two steps are necessary. In a first step, both parties need to verify their trustworthiness mutually and exchange some initial cryptographic credentials. This can e.g. be achieved by means of an certificate environment based on a public key infrastructure (PKI) and only has to be done once for each time an ED joins a new network hosted by an AP. In a second step, which is executed every time a new communication session is started (e.g. for each new URLLC service), both devices mutually authenticate each other and prove their access rights, which can be done in a handshake like manner based on the initially exchanged information. During this stage, they will additionally generate common keying material which is used in order to secure the message exchange within the URLLC session (note that also updating of the keying material has to be considered during the URLLC session). Fig. \ref{plug+trust_logic} shows an overview of the respective procedures.


\begin{figure}[b]
\begin{center}
\includegraphics[width=4.2cm]{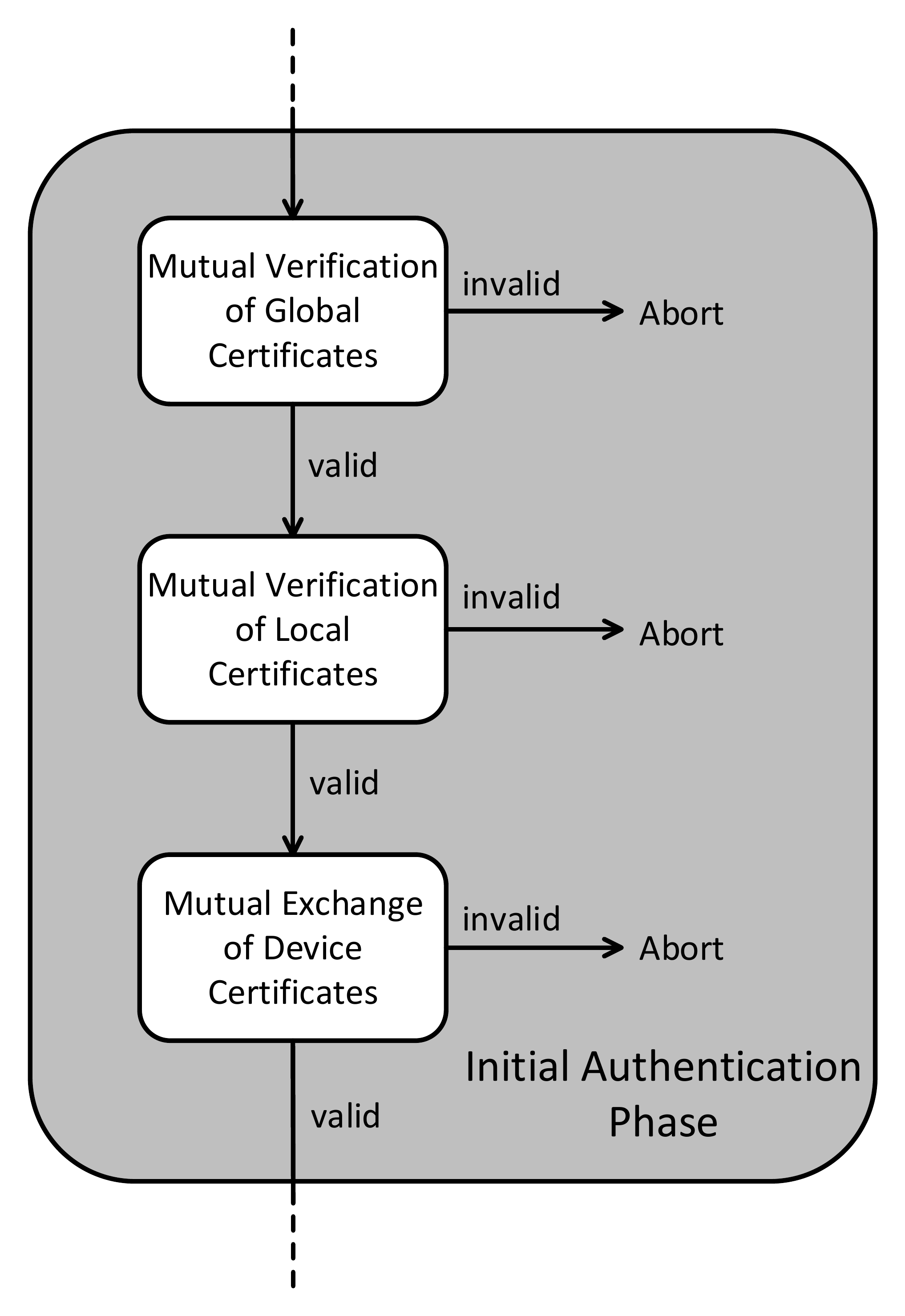}    
\caption{Initial Authentication Phase} 
\label{plug+trust_initial}
\end{center}
\end{figure}

\subsubsection{Initial Authentication Phase}
\begin{figure}[t]
\begin{center}
\includegraphics[width=8.4cm]{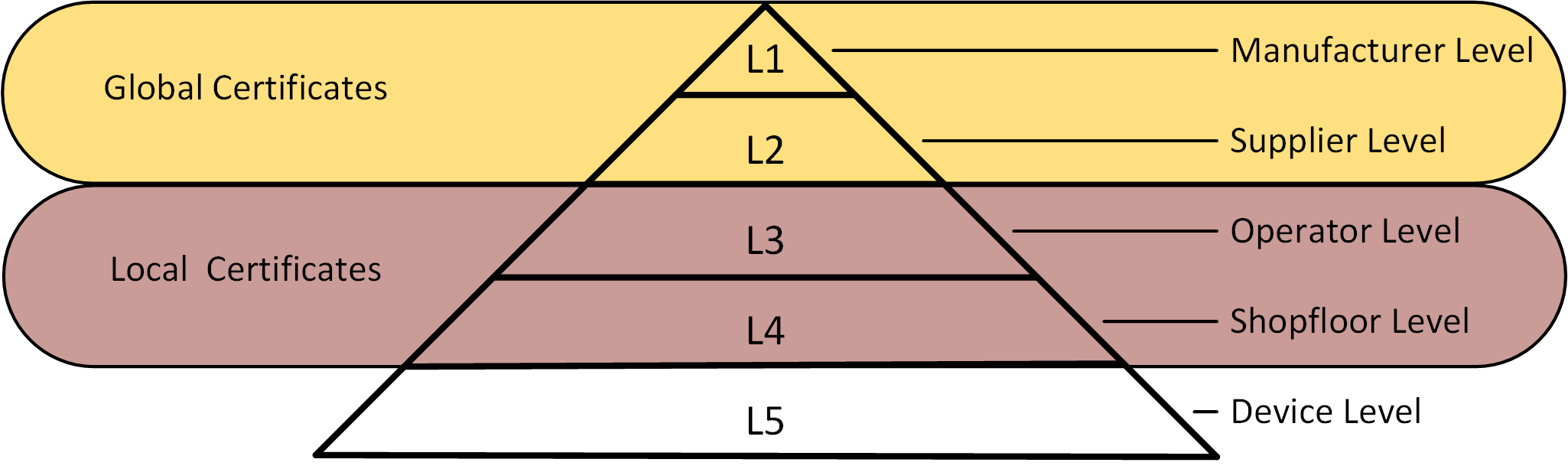}    
\caption{Certificate Authority exemplary hierarchy} 
\label{ca_level}
\end{center}
\end{figure}

Fig. \ref{plug+trust_initial} shows the procedure of an initial connection within a link. This must be carried out each time a new device is added to the network. For that purpose, we propose to make use of a certificate environment which includes different levels of certificate authorities (CA). These may offer global certificates (e.g. specific for the manufacturer of the network equipment or of the SE deployed) as well as local certificates (e.g. operator or plant specific). Thereby, the global certificates enable verification of the respective manufacturer of a network device over the whole device life cycle, whereas local certificates enable the verification of the access rights of a network device to specific networks of an operator or within a plant. More information on this concept, as well as how to load certificates to the network devices initially, can be e.g. found in [\cite{Mackenthun.2017}]. An example of a CA hierarchy including different levels is shown in Fig. \ref{ca_level}. After setting up the physical connection, the two devices (AP and ED) will check their global certificates mutually in order to make sure that no third party devices are involved (which could be already compromised, e.g. due to product piracy). This must be possible any time for a network device (even when not installed in a network) in order to ensure lifecycle security. In case of a successful verification of the global certificates, the next level of certificates is checked which provides access rights e.g. for a specific operator and factory (local certificates). This means, that network devices which are equipped with these certificates could by default not be operated by a third party (e.g. an attacker) and vice versa (third party devices will not get access to the dedicated network). In a final step within the initial authentication phase, both devices need to exchange their own device specific certificates in order to make use of cryptographic techniques introduced within the next section.

\subsubsection{URLLC Session Handshake}
\begin{figure}[b]
\begin{center}
\includegraphics[width=6.3cm]{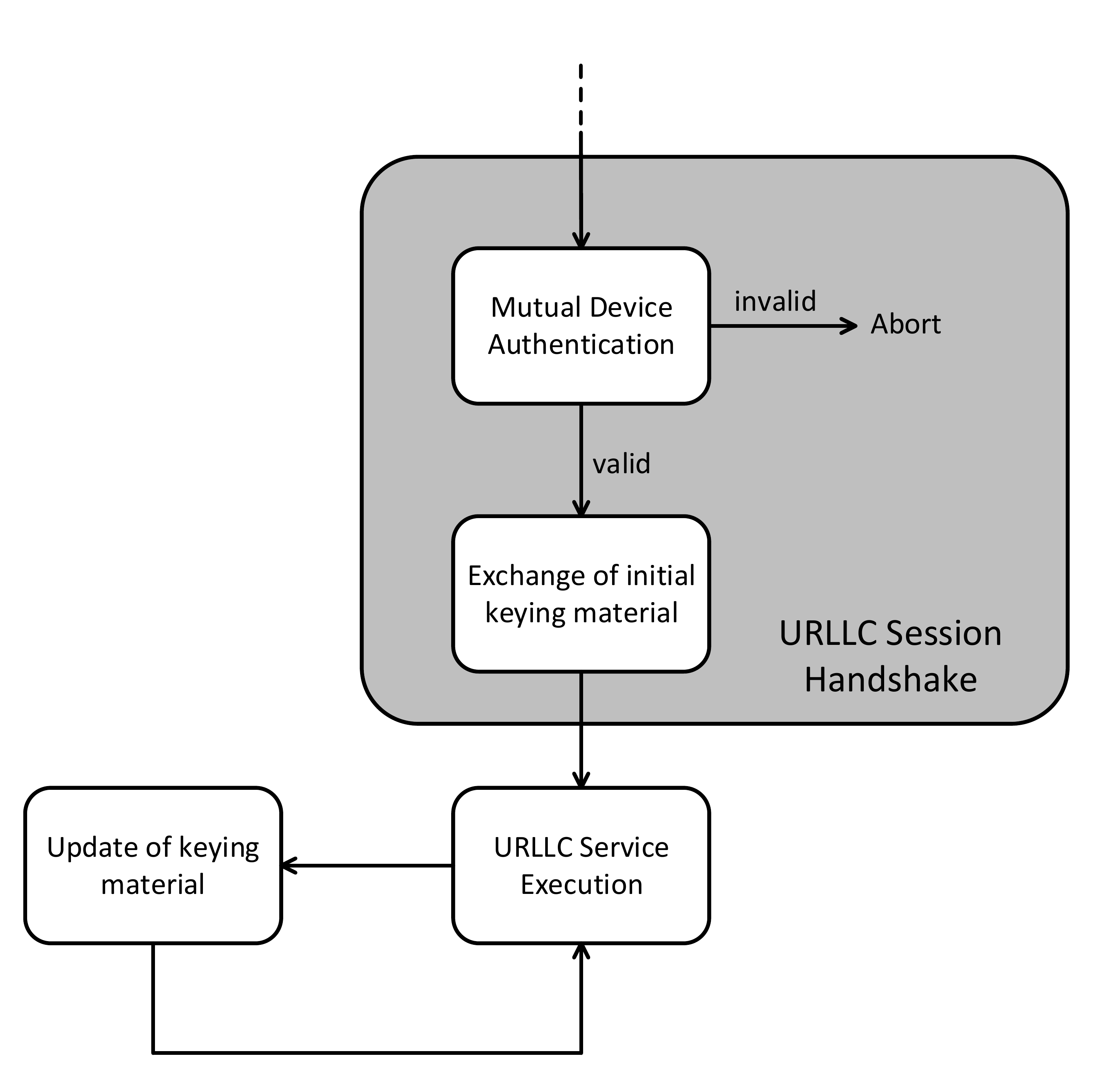}    
\caption{URLLC Session Handshake} 
\label{plug+trust_handshake}
\end{center}
\end{figure}

In order to set up an URLLC session, network devices must authenticate mutually first (see Fig. \ref{plug+trust_handshake}). This means, that they need to prove the possession of the private key matching to the public key which was exchanged within the device specific certificate in the previous step. For this, e.g. a simple mutual challenge-response scheme including random nonce values can be used (such as e.g. within a TLS 1.3 handshake [\cite{IETF2018}]). After a successful authentication of both devices, they can agree on initial keying material (the usage of this is introduced within the next section). Both, mutual authentication and negotiation of keying material can be achieved by making use of the public keys cryptography techniques such as signing and verifying and Diffie-Hellman Key Exchange (DHKE).

\subsection{Secure URLLC Channel}
\begin{figure}[t]
\begin{center}
\includegraphics[width=8.4cm]{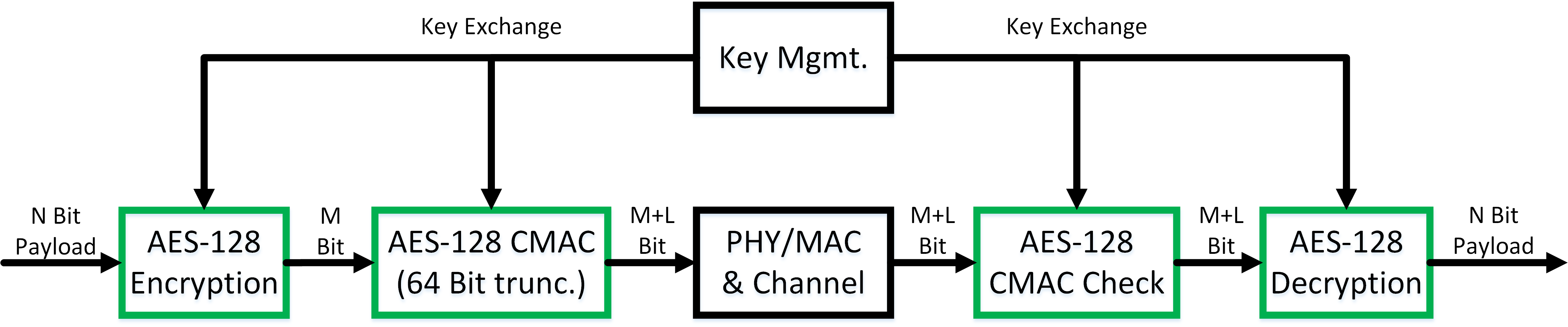}    
\caption{Conventional Security Processing Blocks} 
\label{conventional_sec}
\end{center}
\end{figure}

In this section we introduce the security components that are required in order to ensure secure URLLC data transmission. These are encryption/decryption blocks in order to ensure message confidentiality and message authentication blocks such as MAC tagging and MAC checking. Fig. \ref{conventional_sec} shows the transmitter and receiver security processing blocks using AES-128 as a cipher suite and a $64$ Bit truncated CMAC for message authentication.

\subsubsection{Encryption \& Key Management}
\begin{figure}[b]
\begin{center}
\includegraphics[width=8.4cm]{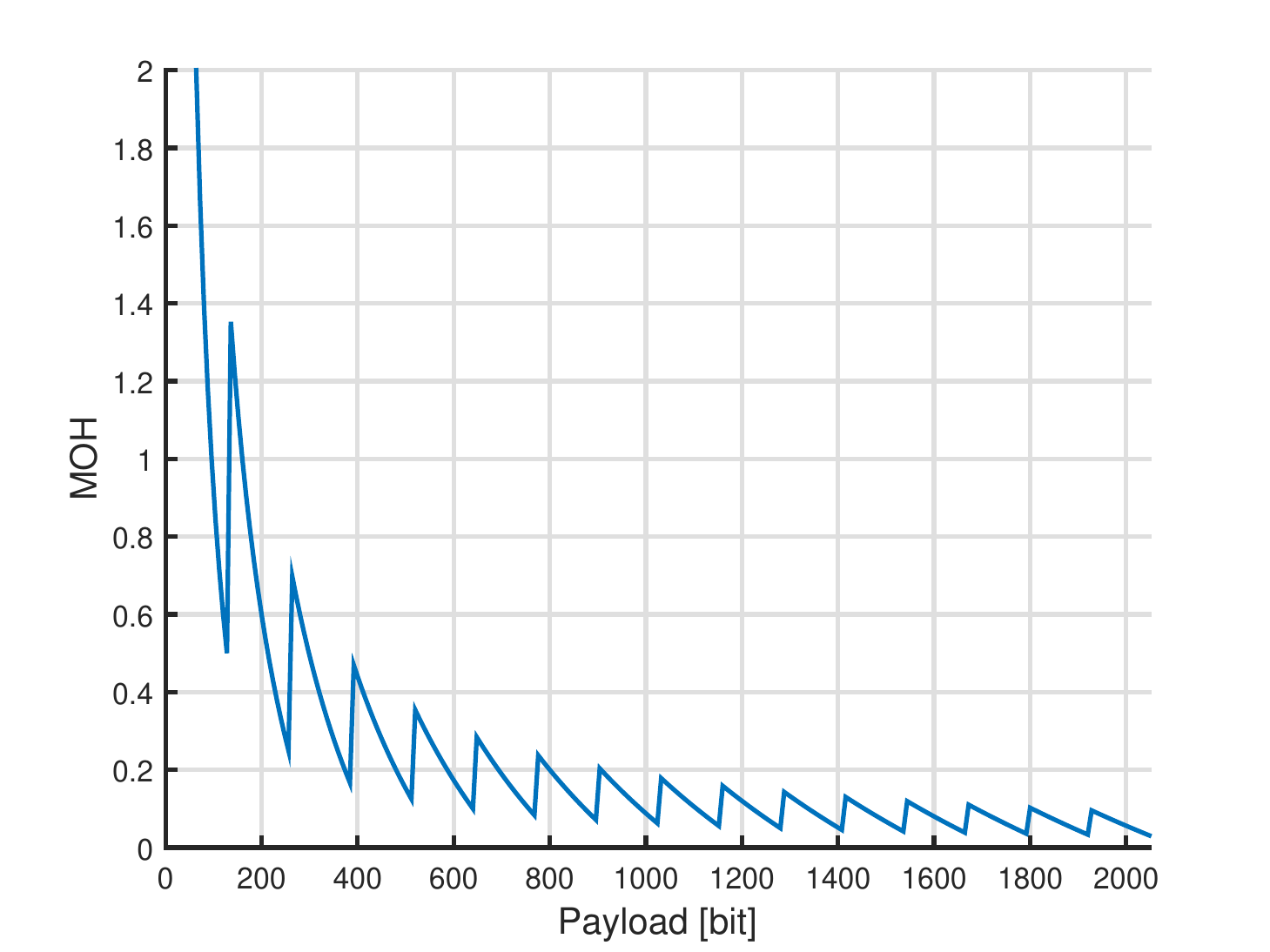}    
\caption{Security overhead due to zero-padding and CMAC} 
\label{sec_oh}
\end{center}
\end{figure}

Block ciphers such as the AES-128 cipher suite take a block of 128 Bit of plaintext data and a 128 Bit symmetric key as input and yield a block of the same size as ciphertext. This comes with the drawback, that URLLC message payload usually needs to be padded with zeros, if the respective application uses other payload sizes than multiples of $128$ Bit. Therefore, even the design of the served control applications will influence the efficiency of the overall system as shown in Fig. \ref{sec_oh}. Another issue of block ciphers is their mode of operation. Typically, information from previous blocks, such as e.g. the ciphertext in case of CBC (Chiper Block Chaining) mode, is fed to the encryption process of the current block. This increases the security, because an potential attacker needs knowledge of all previous ciphertexts, but also comes at the price, that in case of transmission errors the decryption process at the receiver side is delayed. Therefore, an appropriate mode is the CTR (counter) mode for which the encryption and decryption process only depends on the current plaintext/ciphertext and a combined nonce/counter value. This value is similar to the initialisation vector used e.g. in CBC mode, but due to the counter included, ciphertext repetition is prevented. Another drawback of cipher suites is, that keypairs need to be updated regularly in order to ensure a high degree of security such as e.g. forward secrecy. Conventional key management solutions such as, e.g. DHKE, come with the cost of an additional control signalling overhead, which rises with the number of connected devices. Therefore, in section \ref{SKG} an approach based on PHYSEC is introduced in order to reduce the amount of additional signalling overhead.

\begin{figure*}[b]
\begin{center}
\includegraphics[width=\textwidth]{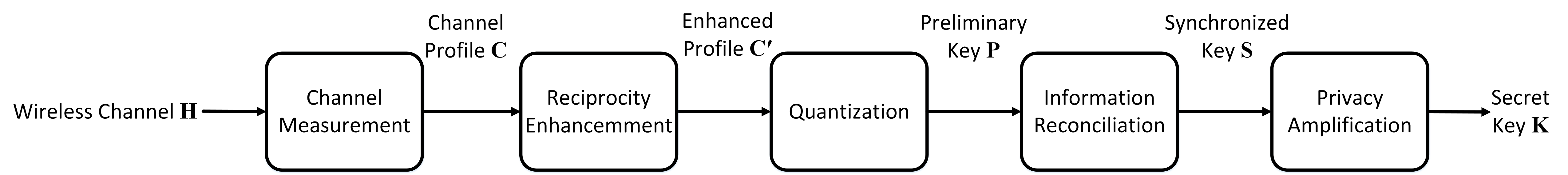}    
\caption{Key Generation Procedure}
\label{key_generation_steps}
\end{center}
\end{figure*}

\subsubsection{Message Authenticity and Integrity}
During data transmission, each message should have a message authentication code (MAC) attached to it in order to ensure message authenticity and integrity. Similar to the encryption process, a given plaintext and a symmetric key are fed as an input to the MAC function and the tag is calculated based on that as an ouput either by applying a block cipher algorithm (such as AES-128) or a secure hash algorithm (e.g. SHA-2 or SHA-3). After reception of the message with the associated MAC by a receiver, the same calculations are applied and the receiver checks if the MAC check holds true (both MAC tags are the same). Though this approach can be assumed as cryptographically secure if the key size is large enough (for AES-128 based CMAC a maximum truncation to 64 Bit is recommended [\cite{RFC_CMAC.2006}] and for hash based MAC (HMAC) a minimum MAC size of 80 Bit is recommended [\cite{RFC_HMAC.1997}]), it requires a non negligible amount of resources by introducing additional overhead to transmitted messages. Additionally, only the message payload can be protected as an attacker might perform attacks such as MAC (Media Access Control) address spoofing, or record a message and replay it after a while (if e.g. counters are used in order to prohibit that, additional overhead is added as well).
Fig. \ref{sec_oh} shows the message overhead (MOH) for a given application payload of $N$ Bit that is added by zero padding due to AES-128 block ciphering and message authentication based on a $L$ Bit MAC tag and calculated as

\begin{equation}
\rm MOH=\frac{\left\lceil \frac{N}{128} \right\rceil \cdot 128 + L}{N}-1.
\end{equation}
In case of e.g. a payload size of $400$ Bit, which is a valid assumption for the payload size of URLLC messages according to [\cite{Bockelmann2017}] and a CMAC size of $64$ Bit, the MOH is $44 \%$. Within Fig. \ref{sec_oh} the MOH is shown for application payload sizes between $64$ Bit (MOH=$200\%$) and $2000$ Bit (MOH=$5.6\%$). In order to reduce this overhead, we introduce an alternative solution based on PHYSEC in section \ref{message_authentication}.

\section{Enhancements based on Physical Layer Security}
\label{physec_section}
\begin{figure}[t]
\begin{center}
\includegraphics[width=4.2cm]{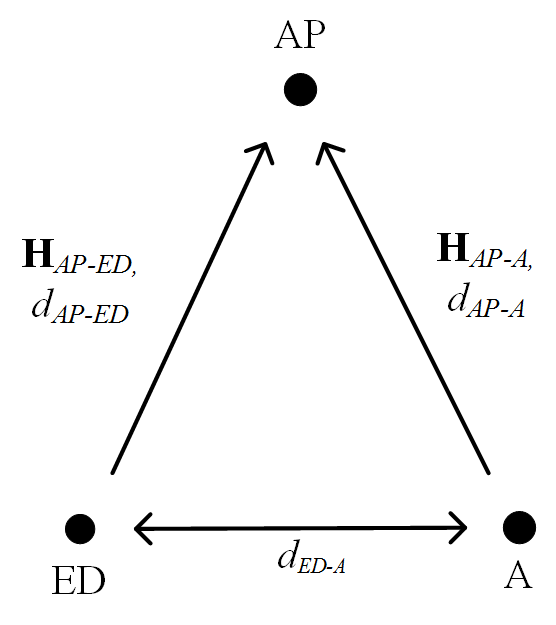}    
\caption{PHYSEC system model} 
\label{physec_system_model}
\end{center}
\end{figure}

Within this section, new security solutions based on PHYSEC are introduced in order to overcome the drawbacks regarding the cryptographic solutions introduced in the previous section. Therefore, a simplified system model is introduced (see Fig. \ref{physec_system_model}) consisting of an AP, an ED and an attacker $A$ and $\boldsymbol{H}^{(k)}_{\mbox{\scriptsize X-Y}}$ denotes the channel matrix $\boldsymbol{H}$ between two users $X$ (Receiver) and $Y$ (Transmitter) at time $k$.

\subsection{Secret Key Generation}
\label{SKG}
In case of Secret Key Generation (SKG), two parties, e.g. the AP and an ED, estimate the channel characteristics they experience mutually for a specific amount of time in order to derive a shared secret which can be used as a session key for ciphering (this approach has e.g. been considered in [\cite{Guillaume.}], [\cite{Zenger.2014}] and [\cite{Wilhelm.2013}]). Let the channel estimated by the ED due to the AP at time $k$ be $\boldsymbol{H}^{(k)}_{\mbox{\scriptsize ED-AP}}$ and vice versa $\boldsymbol{H}^{(k)}_{\mbox{\scriptsize AP-ED}}$. The adversary $A$ eavesdrops and estimates the channel as $\boldsymbol{H}^{(k)}_{\mbox{\scriptsize A-AP}}$ and $\boldsymbol{H}^{(k)}_{\mbox{\scriptsize A-ED}}$ due to the AP and ED respectively. The principle of channel reciprocity indicates that the channel measurements $\boldsymbol{H}^{(k)}_{\mbox{\scriptsize ED-AP}}$ and $\boldsymbol{H}^{(k)}_{\mbox{\scriptsize AP-ED}}$ are equal when they are conducted during the coherence time of the channel and in the absence of errors (e.g. noise). It is now assumed, that $A$ is at minimum located $\frac{\lambda}{2}$ away from both the AP ($d_{\mbox{\scriptsize A-AP}}>\frac{\lambda}{2}$) and the ED ($d_{\mbox{\scriptsize A-ED}}>\frac{\lambda}{2}$) with $\lambda$ being the wavelength of the transmitted wave and $d_{\mbox{\scriptsize A-AP}}$ and $d_{\mbox{\scriptsize A-ED}}$ the distance between the attacker and the AP and ED respectively. If e.g. $\frac{\lambda}{2}\approx 2.5$ cm, corresponding to a frequency of $f=5.8$ GHz, then it is assumed that $\boldsymbol{H}_{\mbox{\scriptsize A-AP}} \ne \boldsymbol{H}_{\mbox{\scriptsize ED-AP}}$ holds, which means that $A$ experiences different channel conditions due to the AP compared to the channel conditions experienced by the ED due to the AP. 

\begin{figure}[t]
\begin{center}
\includegraphics[width=8.4cm]{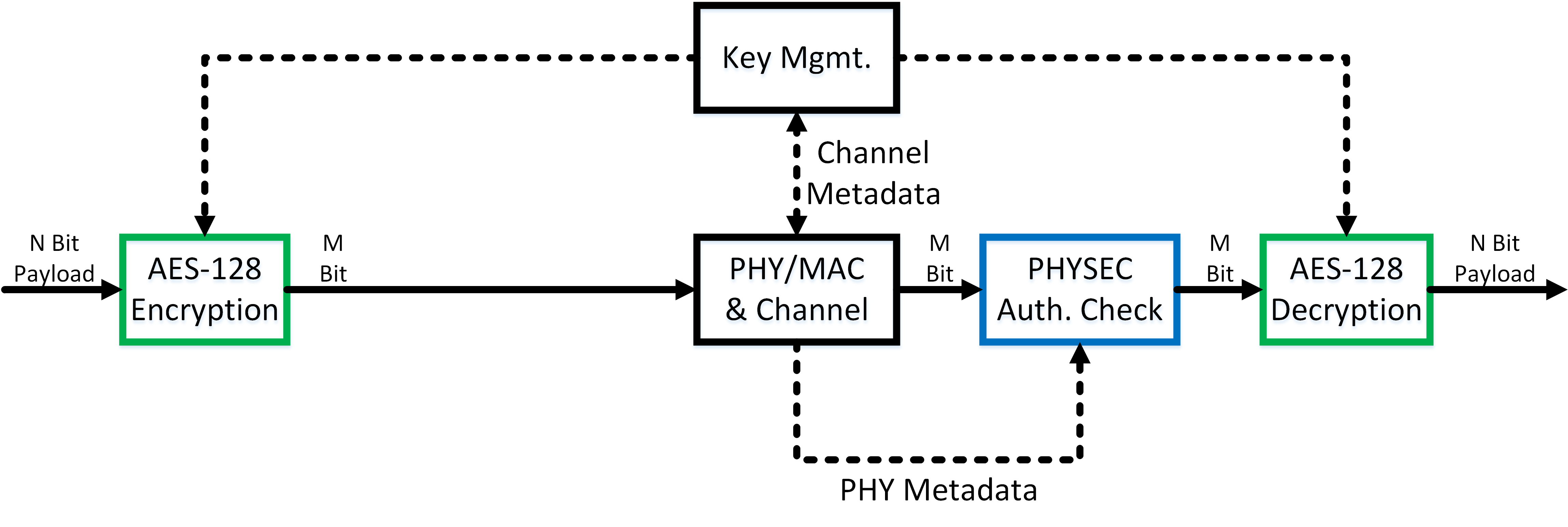}    
\caption{Physical Layer Security Processing Blocks} 
\label{physec}
\end{center}
\end{figure}

\begin{table*}[t]
\caption[caption]{Performance comparison of PHYSEC based key generation approaches}
\label{table_skg}
\centering
\footnotesize
\begin{tabular}{|c|c|c|c|c|}
\hline
Work & Method & Feature & min./max. KGR [Bit/s] & BDR @ min./max. KGR\\
\hline
\hline
[\cite{Guillaume.}] & DCTW & RSS (Wifi) & $0.01$ / $15.45$ & -/- \\
\hline
[\cite{Patwari.2010}] & HRUBE & RSS (802.15.4) & $3$ / $22$ & $0.04\%/2.2\%$ \\
\hline
[\cite{Ambekar.2014}] & Rec. Enhancement & RSS (Wifi) & $3.2$ / $3.41$ & $6.21\%/4.5\%$ \\
\hline
[\cite{Premnath.2013}] & ASBG & RSS (Wifi/802.15.4) & $0.08$ / $0.38$ & $1\%/23\%$ \\
\hline
[\cite{Jana.2009}] & Multibit-ASBG & RSS (Wifi) & $0$ / $0.9$ & $4\%/36\%$ \\
\hline
\end{tabular}
\end{table*}

To now negotiate a secret key of length $L={\mbox{\it{MN}}}$ Bits only known to the AP and ED, several steps need to be executed as shown in Fig. \ref{key_generation_steps}. A common approach is to let the AP and ED each estimate their channel mutually over a period of $M$ measurements, e.g. by means of channel probing. This will yield to channel profiles
$\boldsymbol{C}=(\boldsymbol{H}^{(1)}, \boldsymbol{H}^{(2)},\ldots, \boldsymbol{H}^{(M)})$. 
To enhance the reciprocity of the channel, different approaches, such as e.g. kalman filtering [\cite{Ambekar.2014}] in order to decrease the impact of noise in channel estimations can be applied. In the next step, the enhanced channel profiles $\boldsymbol{C'}_{\mbox{\scriptsize AP-ED}}$ and $\boldsymbol{C'}_{\mbox{\scriptsize ED-AP}}$ derived by both, the AP and ED, are quantized to obtain preliminary keys $\boldsymbol{P}=(P^{(1)}, P^{(2)},\ldots, P^{(L)})$ (see e.g. [\cite{Jana.2009}], [\cite{Patwari.2010}], [\cite{Premnath.2013}]). In general, $N$ Bits of the preliminary key are derived from each channel measurement $\boldsymbol{H}^{(k)}$ ($k=1,\ldots,M$). Due to remaining errors in the channel profiles, errors in the quantization stage occur and disagreement of Bits in preliminary keys exists, i.e. $\boldsymbol{P}_{\mbox{\scriptsize AP-ED}} \ne \ \boldsymbol{P}_{\mbox{\scriptsize ED-AP}}$. These errors are detected and corrected in the information reconciliation stage by means of error correction coding, e.g. turbo codes [\cite{Ambekar.2012b}], yielding the synchronized key $\boldsymbol{S}=\boldsymbol{S}_{\mbox{\scriptsize AP-ED}}=\boldsymbol{S}_{\mbox{\scriptsize ED-AP}}$. Due to parity information exchange between the AP and ED to match their keys during the reconciliation stage, an attacker can use these information to gain partial knowledge of the key. To reduce this effect, as well as to enhance the entropy of the key, privacy amplification is utilized. A common approach here is to make use cryptographic hashing algorithms (e.g. SHA-2, SHA-3). Finally, the AP and ED both share the secret key $\boldsymbol{K}=\boldsymbol{K}_{\mbox{\scriptsize AP-ED}}=\boldsymbol{K}_{\mbox{\scriptsize ED-AP}}$. The performance of different key generation approaches is mainly given by two metrics. The key generation rate (KGR) which is the effictive rate of generated bits of the symmetric key per second. The other one is the bit disagreement rate, which denotes the amount of disagreeing bits between the ED and AP before the Information Reconcilliation stage.

\subsubsection{Performance Comparison}
Tab. \ref{table_skg} shows the performance of different existing SKG approaches. Most approaches consider the Quantization stage in order to improve the KGR, e.g. by extracting multiple Bits ($N>1$) from one channel measurement. Whereas [\cite{Ambekar.2014}] proposes different Reciprocity Enhancement approaches in order to reduce mismatching Bits due to errors in channel measurements between two users. Nevertheless which approach is used, the KGR mainly depends on the temporal decorrelation of the wireless channel and therefore is highly depending on the given scenario. If the temporal channel variations are very low, e.g. in a static scenario, it is not feasible to use this method for generation of initial keying material. But due to the low complexity of the schemes, they can be used to update e.g. encryption keys periodically after the initial key exchange (e.g. using DHKE) in order to improve the security level and keep the additional security control signaling overhead low. This is also shown in Fig. \ref{physec} where only partial information between two parties is exchanged (e.g. the parity bits, denoted by dashed line) compared to a full key exchange in case of Fig. \ref{conventional_sec} denoted by solid lines. 

\begin{table*}[t]
\caption[caption]{Performance comparison of PHYSEC based message authentication approaches}
\label{table_message_authentication}
\centering
\footnotesize
\begin{tabular}{|c|c|c|c|c|}
\hline
Work & Detection Method & Feature & max. $P_{\mbox{\tiny D}}$ & min. $P_{\mbox{\tiny FA}}$\\
\hline
\hline
[\cite{Gulati.2013}] & GMM & Antenna Mode & $99.9\%$ & $0.4\%$ \\
\hline
[\cite{Pei.2014}] & SVM, LFDA & RSSI, TOA, CC & $95\%$ & $1\%$ \\
\hline
[\cite{Tugnait.2010}] & NPHT & CIR (TD) & $99.9\%$ & $1\%$ \\
\hline
[\cite{Xiao.2008}] & NPHT & CIR (FD) & $99\%$ & $1\%$ \\
\hline
[\cite{Xiao.2017}] & Q-Learn., NE GT & CIR (FD) & 98\% & 0.1\% \\
\hline
[\cite{Weinand.2017}] & GMM & CIR (FD) & $99.96\%$ & $0.1\%$ \\
\hline
\end{tabular}
\end{table*}

\subsection{PHYSEC based Message Authentication and Integrity Checking}
\label{message_authentication}
\begin{figure}[h]
\centering
\includegraphics[width=0.15\textwidth]{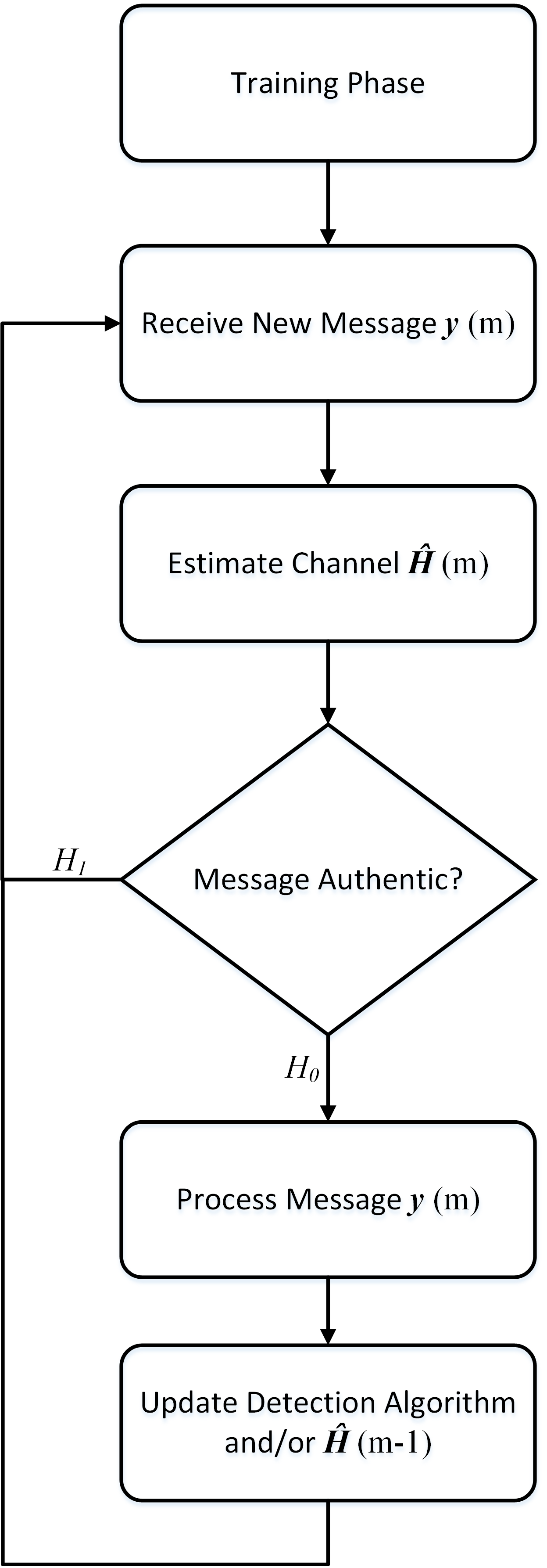}
\caption{Message Authentication based on Physical Layer Metadata}
\label{ML_flow_diag}
\end{figure}

An alternative approach in order to guarantee message authenticity and integrity is to take non-cryptographic information into account such as protocol metadata. These can either be extracted from protocols on the logical level such as frame counters or traffic patterns, or they can be extracted from physical layer protocols such as received signal strength or channel estimation data. In case of the schemes based on higher layer protocols, it is very easy for an attacker to spoof messages if he knows the used protocol and the traffic pattern of a legitimate user (e.g. the radio resources that are allocated for that user). If the receiver analyses the physical characteristics of the received signal which are introduced by the transmitter such as e.g. radio fingerprints or clock skew, then it is already challenging for an attacker to spoof these characteristics by using default hardware (off-the-shelf). On the other hand, this also requires an additional effort for the legitimate receiver in order to detect these characteristics and by this increases the receiver complexity and cost. Therefore, another approach is to use characteristics of the wireless channel such as channel state information, received signal strength (RSS) or carrier frequency offsets (CFO) due to doppler shifts in order to authenticate the transmitter of a message and its integrity. The main advantage that can be used here is the spatial decorrelation property of the wireless channel, which denotes that channel characteristics are varying based on the location of a transceiver. This holds already true for small variations in the spatial domain which are in the order of the wavelength of the transmitted signal. 
In general, this approach contains two stages. In the first stage, which is a kind of training phase, physical layer metadata is extracted initially and stored by the intended receiver. If we assume that an ED wants to transmit authenticated messages to the AP and the channel estimates are used as a feature, the AP will estimate the channel as $\boldsymbol{\hat{H}}_{\mbox{\scriptsize AP-ED}}{(k)}$ during the training phase for each of the $T$ received messages $\boldsymbol{y}_{\mbox{\scriptsize AP}}{(k)}, k=0,\ldots,T-1$. In this stage we also assume, that the acquired data is labeled. This means that the messages $\boldsymbol{x}_{\mbox{\scriptsize ED}}(k)$ sent by the ED contain cryptographic information from higher layers (e.g. certificates or digital signatures). In the second stage, the ED transmits messages without any cryptographic information attached and the AP uses the knowledge from the training phase in order to make a decision on these further received messages $\boldsymbol{y}_{\mbox{\scriptsize AP}}{(m)}$ at time $m\ge T$ with respect to the authenticity and integrity. If the AP estimates the channel now as $\boldsymbol{\hat{H}}{(m)}$, there are two hypothesis, either
\begin{equation} \label{eq1}
\begin{split}
\mathcal{H}_0&: \boldsymbol{y}_{\mbox{\scriptsize AP}}{(m)}, \ \boldsymbol{x}(m) \ \rm sent\ by\ ED,\ or\ \\
\mathcal{H}_1&: \boldsymbol{y}_{\mbox{\scriptsize AP}}{(m)}, \ \boldsymbol{x}(m) \ \rm not\ sent\ by\ ED,\
\end{split}
\end{equation}
where $\boldsymbol{x}(m)$ is the originally sent message. In case of $\mathcal{H}_0$
\begin{equation}
\boldsymbol{\hat{H}}{(m)}=\boldsymbol{H}_{\mbox{\scriptsize AP-ED}}{(m)}+\boldsymbol{\epsilon}_{\mbox{\scriptsize AP-ED}}{(m)},
\label{eq:}
\end{equation}  
and in case of hypothesis $\mathcal{H}_1$
\begin{equation}
\boldsymbol{\hat{H}}{(m)}\ne\boldsymbol{H}_{\mbox{\scriptsize AP-ED}}{(m)}+\boldsymbol{\epsilon}_{\mbox{\scriptsize AP-ED}}{(m)},
\label{eq:}
\end{equation}
where $\boldsymbol{\epsilon}_{\mbox{\scriptsize X-Y}}(m)$ is the estimation error (e.g. due to receiver noise) of the true channel $\boldsymbol{H}_{\mbox{\scriptsize X-Y}}(m)$ between users $X$ and $Y$. Due to temporal variations in the channel $\boldsymbol{H}_{\mbox{\scriptsize AP-ED}}$, these need to be somehow considered in the second stage. Otherwise it is not possible to distinguish between temporal channel variations and messages introduced by an attacker anymore at some point. There are two possibilities in order to achieve that goal, either the detection algorithm needs to be updated regularly, or instead of this, the difference of the current channel estimate to the previous one 
\begin{equation}
\Delta\boldsymbol{\hat{H}}{(m)}=\boldsymbol{\hat{H}}{(m)}-\boldsymbol{\hat{H}}{(m-1)}
\label{eq:}
\end{equation}
is used as input data for the detection algorithm. In the first case, the update interval should be below the channel coherence time $T_c$ of the $\boldsymbol{H}_{\mbox{\scriptsize AP-ED}}$ channel in order to catch up with these variations. Whereas in both cases, the temporal difference between two subsequent channel estimations should always be below $T_c$. Fig. \ref{ML_flow_diag} shows all steps of the general procedure described above.
As detection algorithm, simple statistical test such as Neyman Pearson Hypothesis Testing (NPHT), e.g. used in \cite{Xiao.2008}, or more sophisticated machine learning algorithms such Gaussian Mixture Models introduced in \cite{Gulati.2013} can be utilized. Every detection algorithm has basically two main performance indicators, the detection probability $P_{\mbox{\scriptsize D}}$ and the false alarm rate $P_{\mbox{\scriptsize FA}}$. In our case, the detection probability 
\begin{equation}
P_{\mbox{\scriptsize D}}=p(\mathcal{H}_1|\boldsymbol{\hat{H}}{(m)} \mathrm{\ due\ to\ } \boldsymbol{x}_{\mbox{\scriptsize A}})
\label{eq:}
\end{equation}
denotes the probability of detecting the attacker $A$ as the transmitter of $\boldsymbol{y}_{\mbox{\scriptsize AP}}{(m)}$ under the condition that it was truly sent by the attacker and the false alarm rate
\begin{equation}
P_{\mbox{\scriptsize FA}}=p(\mathcal{H}_1|\boldsymbol{\hat{H}}{(m)} \mathrm{\ due\ to\ } \boldsymbol{x}_{\mbox{\scriptsize ED}})
\label{eq:}
\end{equation}
denotes the probability of detecting the attacker as the transmitter of $\boldsymbol{y}_{\mbox{\scriptsize AP}}{(m)}$ under the condition that it was truly sent by the ED. Both performance metrics need to be considered jointly, as they can not be optimized independently. The relation of both can be shown in a receiver operating characteristic (ROC) curve.


\subsubsection{Performance Comparison}
A performance comparison of different PHYSEC based message authentication approaches is given in Tab. \ref{table_message_authentication}, where the detection methods, used features, maximum achieved detection probability and minimum false alarm rate are considered. Though many of the approaches seem to have quite promising performances (e.g. up to $99.9\%$ detection probability and $0.1\%$ false alarm rate), there is still some improvement in reliability needed in order to serve for the requirements of URLLC services. One possible approach for that is to use additional features within the detection process such as RSSI or CFO simultaneously. Further, the tradeoff between complexity and performance of the detection method is an issue that needs to be addressed in future work, as better performing solutions usually come with a cost. If this can be achieved, the latency and resource efficiency can be increased by omitting the application of MAC tags for message authentication and integrity checking and using physical layer meta information for that purpose only (as shown by dashed line in Fig. \ref{physec}).

\section{Conclusion and Future Work}
\label{CONC}
Within this work we have proposed security solutions for wireless networks in industrial applications, especially with respect to the needs of URLLC traffic. Especially PHYSEC might be a promising approach in order to overcome drawbacks of conventional security solutions in terms of resource efficiency. One essential part of the security design is the user authentication procedure and deployment of certificate environments, as this deals as a root of trust for all other security features such as message encryption and authentication (even if PHYSEC approaches are applied). Therefore, we especially emphasise this point for the design of new systems or whole ecosystems.

\begin{ack}
This work has been supported by the Federal Ministry of Education and Research of the Federal Republic of Germany (BMBF) in the framework of the project 16KIS0267 HiFlecs. The authors alone are responsible for the content of the paper. 
\end{ack}

\bibliography{arch_references}             

\begin{thebibliography}{27}
\providecommand{\natexlab}[1]{#1}
\providecommand{\url}[1]{\texttt{#1}}
\providecommand{\urlprefix}{URL }
\expandafter\ifx\csname urlstyle\endcsname\relax
  \providecommand{\doi}[1]{doi:\discretionary{}{}{}#1}\else
  \providecommand{\doi}{doi:\discretionary{}{}{}\begingroup
  \urlstyle{rm}\Url}\fi

\bibitem[{Ambekar et~al.(2012)Ambekar, Hassan, and Schotten}]{Ambekar.2012b}
Ambekar, A., Hassan, M., and Schotten, H.D. (2012).
\newblock Improving channel reciprocity for effective key management systems.
\newblock In \emph{International Symposium on Signals, Systems and Electronics
  (ISSSE), Potsdam, Germany}.
\newblock \doi{10.1109/ISSSE.2012.6374318}.

\bibitem[{Ambekar and Schotten(2014)}]{Ambekar.2014}
Ambekar, A. and Schotten, H.D. (2014).
\newblock Enhancing channel reciprocity for effective key management in
  wireless ad-hoc networks.
\newblock In \emph{IEEE Vehicular Technology Conference (VTC Spring), Seoul,
  Korea}.
\newblock \doi{10.1109/VTCSpring.2014.7022913}.

\bibitem[{{Bluetooth SIG}(2016)}]{bluetooth}
{Bluetooth SIG} (2016).
\newblock {Bluetooth Core Specification v5.0}.

\bibitem[{Bockelmann et~al.(2017)Bockelmann, Dekorsy, Gnad, Rauchhaupt,
  Neumann, Block, Meier, Rust, Paul, Mackenthun et~al.}]{Bockelmann2017}
Bockelmann, C., Dekorsy, A., Gnad, A., Rauchhaupt, L., Neumann, A., Block, D.,
  Meier, U., Rust, J., Paul, S., Mackenthun, F., et~al. (2017).
\newblock Hiflecs: Innovative technologies for low-latency wireless closed-loop
  industrial automation systems.
\newblock \emph{22. VDE-ITG-Fachtagung Mobilkommunikation}.

\bibitem[{Guillaume et~al.(2015)Guillaume, Winzer, Czylwik, Zenger, and
  Paar}]{Guillaume.}
Guillaume, R., Winzer, F., Czylwik, A., Zenger, C.T., and Paar, C. (2015).
\newblock Bringing phy-based key generation into the field: An evaluation for
  practical scenarios.
\newblock In \emph{IEEE 82nd Vehicular Technology Conference (VTC Fall)}.
\newblock \doi{10.1109/VTCFall.2015.7390857}.

\bibitem[{Gulati et~al.(2013)Gulati, Greenstadt, Dandekar, and
  Walsh}]{Gulati.2013}
Gulati, N., Greenstadt, R., Dandekar, K.R., and Walsh, J.M. (2013).
\newblock Gmm based semi-supervised learning for channel-based authentication
  scheme.
\newblock In \emph{IEEE Vehicular Technology Conference (VTC Fall)}.
\newblock \doi{10.1109/VTCFall.2013.6692216}.

\bibitem[{IEEE(2005)}]{802_15_1_standard}
IEEE (2005).
\newblock {{IEEE} Standard for Information technology-- Local and metropolitan
  area networks-- Specific requirements-- Part 15.1a: Wireless Medium Access
  Control ({MAC}) and Physical Layer ({PHY}) specifications for Wireless
  Personal Area Networks ({WPAN})}.
\newblock \emph{IEEE Std 802.15.1-2005 (Revision of IEEE Std 802.15.1-2002)}.
\newblock \doi{10.1109/IEEESTD.2005.96290}.

\bibitem[{IEEE(2010)}]{802_1X_standard}
IEEE (2010).
\newblock {IEEE Standard for Local and metropolitan area networks--Port-Based
  Network Access Control}.
\newblock \emph{IEEE Std 802.1X-2010 (Revision of IEEE Std 802.1X-2004)}.
\newblock \doi{10.1109/IEEESTD.2010.5409813}.

\bibitem[{IEEE(2016{\natexlab{a}})}]{802_11_standard}
IEEE (2016{\natexlab{a}}).
\newblock {IEEE Standard for Information technology--Telecommunications and
  information exchange between systems Local and metropolitan area
  networks--Specific requirements - Part 11: Wireless LAN Medium Access Control
  (MAC) and Physical Layer (PHY) Specifications}.
\newblock \emph{IEEE Std 802.11-2016 (Revision of IEEE Std 802.11-2012)}.
\newblock \doi{10.1109/IEEESTD.2016.7786995}.

\bibitem[{IEEE(2016{\natexlab{b}})}]{802_15_4_standard}
IEEE (2016{\natexlab{b}}).
\newblock {IEEE Standard for Low-Rate Wireless Networks}.
\newblock \emph{IEEE Std 802.15.4-2015 (Revision of IEEE Std 802.15.4-2011)}.
\newblock \doi{10.1109/IEEESTD.2016.7460875}.

\bibitem[{IETF(1997)}]{RFC_HMAC.1997}
IETF (1997).
\newblock \emph{RFC 2104, HMAC: Keyed-Hashing for Message Authentication}.

\bibitem[{IETF(2006)}]{RFC_CMAC.2006}
IETF (2006).
\newblock \emph{RFC 4493, The AES-CMAC Algorithm}.

\bibitem[{IETF(2018)}]{IETF2018}
IETF (2018).
\newblock \emph{The Transport Layer Security (TLS) Protocol Version 1.3}.

\bibitem[{ITU-R(2015)}]{ITU-R2015}
ITU-R (2015).
\newblock Imt vision--framework and overall objectives of the future
  development of imt for 2020 and beyond.
\newblock \emph{Recommendation ITU-R M.2083-0}.

\bibitem[{Jana et~al.(2009)Jana, Premnath, Clark, Kasera, Patwari, and
  Krishnamurthy}]{Jana.2009}
Jana, S., Premnath, S.N., Clark, M., Kasera, S.K., Patwari, N., and
  Krishnamurthy, S.V. (2009).
\newblock On the effectiveness of secret key extraction from wireless signal
  strength in real environments.
\newblock ACM, New York, NY.
\newblock \doi{10.1145/1614320.1614356}.

\bibitem[{Mackenthun and Berg(2017)}]{Mackenthun.2017}
Mackenthun, F. and Berg, J. (2017).
\newblock Secure machine-to-machine communication.
\newblock In \emph{SmartCard Workshop, Darmstadt, February 2017}.

\bibitem[{NIST(2001)}]{AES_standard}
NIST (2001).
\newblock {Advanced Encryption Standard (AES)}.

\bibitem[{Patwari et~al.(2010)Patwari, Croft, Jana, and Kasera}]{Patwari.2010}
Patwari, N., Croft, J., Jana, S., and Kasera, S.K. (2010).
\newblock High-rate uncorrelated bit extraction for shared secret key
  generation from channel measurements.
\newblock \emph{IEEE Transactions on Mobile Computing}, 9(1), 17--30.
\newblock \doi{10.1109/TMC.2009.88}.

\bibitem[{Pei et~al.(2014)Pei, Zhang, Shen, and Mark}]{Pei.2014}
Pei, C., Zhang, N., Shen, X.S., and Mark, J.W. (2014).
\newblock Channel-based physical layer authentication.
\newblock In \emph{IEEE Global Communications Conference (GLOBECOM)}.
\newblock \doi{10.1109/GLOCOM.2014.7037452}.

\bibitem[{Premnath et~al.(2013)Premnath, Jana, Croft, Gowda, Clark, Kasera,
  Patwari, and Krishnamurthy}]{Premnath.2013}
Premnath, S.N., Jana, S., Croft, J., Gowda, P.L., Clark, M., Kasera, S.K.,
  Patwari, N., and Krishnamurthy, S.V. (2013).
\newblock Secret key extraction from wireless signal strength in real
  environments.
\newblock \emph{IEEE Transactions on Mobile Computing}, 12(5), 917--930.
\newblock \doi{10.1109/TMC.2012.63}.

\bibitem[{Tugnait and Kim(2010)}]{Tugnait.2010}
Tugnait, J.K. and Kim, H. (2010).
\newblock A channel-based hypothesis testing approach to enhance user
  authentication in wireless networks.
\newblock In \emph{International Conference on COMmunication Systems and
  NETworks (COMSNETS 2010)}.
\newblock \doi{10.1109/COMSNETS.2010.5432018}.

\bibitem[{Weinand et~al.(2017)Weinand, Karrenbauer, Sattiraju, and
  Schotten}]{Weinand.2017}
Weinand, A., Karrenbauer, M., Sattiraju, R., and Schotten, H. (2017).
\newblock Application of machine learning for channel based message
  authentication in mission critical machine type communication.
\newblock In \emph{European Wireless 2017; 23th European Wireless Conference},
  1--5.

\bibitem[{Wilhelm et~al.(2013)Wilhelm, Martinovic, and Schmitt}]{Wilhelm.2013}
Wilhelm, M., Martinovic, I., and Schmitt, J.B. (2013).
\newblock Secure key generation in sensor networks based on frequency-selective
  channels.
\newblock \emph{IEEE Journal on Selected Areas in Communications}, 31(9),
  1779--1790.
\newblock \doi{10.1109/JSAC.2013.130911}.

\bibitem[{Xiao et~al.(2017)Xiao, Chen, Han, Zhuang, and Sun}]{Xiao.2017}
Xiao, L., Chen, T., Han, G., Zhuang, W., and Sun, L. (2017).
\newblock Game theoretic study on channel-based authentication in mimo systems.
\newblock \emph{IEEE Transactions on Vehicular Technology}, PP(99), 1--1.
\newblock \doi{10.1109/TVT.2017.2652484}.

\bibitem[{Xiao et~al.(2008)Xiao, Greenstein, Mandayam, and Trappe}]{Xiao.2008}
Xiao, L., Greenstein, L., Mandayam, N., and Trappe, W. (2008).
\newblock Using the physical layer for wireless authentication in time-variant
  channels.
\newblock \emph{IEEE Transactions on Wireless Communications}, 7(7),
  2571--2579.
\newblock \doi{10.1109/TWC.2008.070194}.

\bibitem[{Zenger et~al.(2014)Zenger, Chur, Posielek, Paar, and
  Wunder}]{Zenger.2014}
Zenger, C.T., Chur, M.J., Posielek, J.F., Paar, C., and Wunder, G. (2014).
\newblock A novel key generating architecture for wireless low-resource
  devices.
\newblock In \emph{International Workshop on Secure Internet of Things (SIoT)}.
\newblock \doi{10.1109/SIoT.2014.7}.

\bibitem[{{Zigbee Alliance}(2012)}]{zigbee}
{Zigbee Alliance} (2012).
\newblock {Zigbee Specification}.

\end{thebibliography}
                                                   







\end{document}